\documentclass[%
 reprint,
 superscriptaddress,
 amsmath,amssymb,
 aps,
 prl,
]{revtex4-2}

\usepackage[T5,T1]{fontenc}

\usepackage[section]{placeins}

\usepackage{mwe}
\usepackage{xcolor}
\definecolor{newred}{RGB}{209,53,43}
\definecolor{newblue}{RGB}{74,125,185}

\usepackage{hyperref}

\hypersetup
{
    colorlinks=true,
    linkcolor={newred},
    urlcolor={newblue},
    filecolor={newred},
    citecolor={newblue},
}

\usepackage{array}
\usepackage{booktabs}
\usepackage{multirow}

\newcommand{\labfig}[1]{\label{fig:#1}}
\newcommand{\reffig}[1]{\hyperref[fig:#1]{Figure}~\ref{fig:#1}\xspace}

\begin{document}
\title{A search for neutrino emission from cores of Active Galactic Nuclei}

% \title{A first indication of neutrino emission from cores of Active Galactic Nuclei}
% \author{IceCube Collaboration}

\begin{abstract}
The sources of the majority of the high-energy astrophysical neutrinos observed with the IceCube neutrino telescope at the South Pole are unknown. So far, only a gamma-ray blazar was {compellingly} associated with the emission of high-energy neutrinos. In addition, several studies suggest that the neutrino emission from the gamma-ray blazar population only {accounts} for a small fraction of the total {astrophysical} neutrino flux. In this work we probe the production of high-energy neutrinos in the cores of Active Galactic Nuclei (AGN), induced by accelerated cosmic rays in the accretion disk region.
We present a likelihood analysis based on eight years of IceCube data, searching for a cumulative neutrino {signal} from three AGN samples created for this work. The neutrino emission is assumed to be proportional to the accretion disk luminosity estimated from the soft X-ray flux. Next to the observed soft X-ray flux, the objects for the three samples have been selected based on their radio emission and infrared color properties.
%For the largest sample in this search, the best fit finds {$105\pm44$ signal-like neutrinos in addition to the expected background of atmospheric and astrophysical neutrinos}, corresponding to a post-trial significance of {2.60}$\sigma$. 
{{For the largest sample in this search, an excess of high-energy neutrino events with respect to {an isotropic background of atmospheric and astrophysical neutrinos} is found, corresponding to a post-trial significance of 2.60$\sigma$.}} 
Assuming a power-law spectrum, the best-fit spectral index is $2.03^{+0.14}_{-0.11}$, consistent with expectations from particle acceleration in astrophysical sources. 
{If interpreted as a genuine signal with the assumptions of a proportionality of X-ray and neutrino fluxes and a model for the sub-threshold flux distribution, this observation implies that at 100~TeV, {{27\% -- 100\%}} of the observed neutrinos arise from particle acceleration in the core of AGN.}
%$96\pm69\%$ of the observed neutrinos arise from particle acceleration in the core of AGN.}
\end{abstract}

% version at the 2021-11-15
\affiliation{III. Physikalisches Institut, RWTH Aachen University, D-52056 Aachen, Germany}
\affiliation{Department of Physics, University of Adelaide, Adelaide, 5005, Australia}
\affiliation{Dept. of Physics and Astronomy, University of Alaska Anchorage, 3211 Providence Dr., Anchorage, AK 99508, USA}
\affiliation{Dept. of Physics, University of Texas at Arlington, 502 Yates St., Science Hall Rm 108, Box 19059, Arlington, TX 76019, USA}
\affiliation{CTSPS, Clark-Atlanta University, Atlanta, GA 30314, USA}
\affiliation{School of Physics and Center for Relativistic Astrophysics, Georgia Institute of Technology, Atlanta, GA 30332, USA}
\affiliation{Dept. of Physics, Southern University, Baton Rouge, LA 70813, USA}
\affiliation{Dept. of Physics, University of California, Berkeley, CA 94720, USA}
\affiliation{Lawrence Berkeley National Laboratory, Berkeley, CA 94720, USA}
\affiliation{Institut f{\"u}r Physik, Humboldt-Universit{\"a}t zu Berlin, D-12489 Berlin, Germany}
\affiliation{Fakult{\"a}t f{\"u}r Physik {\&} Astronomie, Ruhr-Universit{\"a}t Bochum, D-44780 Bochum, Germany}
\affiliation{Universit{\'e} Libre de Bruxelles, Science Faculty CP230, B-1050 Brussels, Belgium}
\affiliation{Vrije Universiteit Brussel (VUB), Dienst ELEM, B-1050 Brussels, Belgium}
\affiliation{Department of Physics and Laboratory for Particle Physics and Cosmology, Harvard University, Cambridge, MA 02138, USA}
\affiliation{Dept. of Physics, Massachusetts Institute of Technology, Cambridge, MA 02139, USA}
\affiliation{Dept. of Physics and Institute for Global Prominent Research, Chiba University, Chiba 263-8522, Japan}
\affiliation{Department of Physics, Loyola University Chicago, Chicago, IL 60660, USA}
\affiliation{Dept. of Physics and Astronomy, University of Canterbury, Private Bag 4800, Christchurch, New Zealand}
\affiliation{Dept. of Physics, University of Maryland, College Park, MD 20742, USA}
\affiliation{Dept. of Astronomy, Ohio State University, Columbus, OH 43210, USA}
\affiliation{Dept. of Physics and Center for Cosmology and Astro-Particle Physics, Ohio State University, Columbus, OH 43210, USA}
\affiliation{Niels Bohr Institute, University of Copenhagen, DK-2100 Copenhagen, Denmark}
\affiliation{Dept. of Physics, TU Dortmund University, D-44221 Dortmund, Germany}
\affiliation{Dept. of Physics and Astronomy, Michigan State University, East Lansing, MI 48824, USA}
\affiliation{Dept. of Physics, University of Alberta, Edmonton, Alberta, Canada T6G 2E1}
\affiliation{Erlangen Centre for Astroparticle Physics, Friedrich-Alexander-Universit{\"a}t Erlangen-N{\"u}rnberg, D-91058 Erlangen, Germany}
\affiliation{Physik-department, Technische Universit{\"a}t M{\"u}nchen, D-85748 Garching, Germany}
\affiliation{D{\'e}partement de physique nucl{\'e}aire et corpusculaire, Universit{\'e} de Gen{\`e}ve, CH-1211 Gen{\`e}ve, Switzerland}
\affiliation{Dept. of Physics and Astronomy, University of Gent, B-9000 Gent, Belgium}
\affiliation{Dept. of Physics and Astronomy, University of California, Irvine, CA 92697, USA}
\affiliation{Karlsruhe Institute of Technology, Institute for Astroparticle Physics, D-76021 Karlsruhe, Germany }
\affiliation{Karlsruhe Institute of Technology, Institute of Experimental Particle Physics, D-76021 Karlsruhe, Germany }
\affiliation{Dept. of Physics, Engineering Physics, and Astronomy, Queen's University, Kingston, ON K7L 3N6, Canada}
\affiliation{Dept. of Physics and Astronomy, University of Kansas, Lawrence, KS 66045, USA}
\affiliation{Department of Physics and Astronomy, UCLA, Los Angeles, CA 90095, USA}
\affiliation{Centre for Cosmology, Particle Physics and Phenomenology - CP3, Universit{\'e} catholique de Louvain, Louvain-la-Neuve, Belgium}
\affiliation{Department of Physics, Mercer University, Macon, GA 31207-0001, USA}
\affiliation{Dept. of Astronomy, University of Wisconsin{\textendash}Madison, Madison, WI 53706, USA}
\affiliation{Dept. of Physics and Wisconsin IceCube Particle Astrophysics Center, University of Wisconsin{\textendash}Madison, Madison, WI 53706, USA}
\affiliation{Institute of Physics, University of Mainz, Staudinger Weg 7, D-55099 Mainz, Germany}
\affiliation{Department of Physics, Marquette University, Milwaukee, WI, 53201, USA}
\affiliation{Institut f{\"u}r Kernphysik, Westf{\"a}lische Wilhelms-Universit{\"a}t M{\"u}nster, D-48149 M{\"u}nster, Germany}
\affiliation{Bartol Research Institute and Dept. of Physics and Astronomy, University of Delaware, Newark, DE 19716, USA}
\affiliation{Dept. of Physics, Yale University, New Haven, CT 06520, USA}
\affiliation{Dept. of Physics, University of Oxford, Parks Road, Oxford OX1 3PU, UK}
\affiliation{Dept. of Physics, Drexel University, 3141 Chestnut Street, Philadelphia, PA 19104, USA}
\affiliation{Physics Department, South Dakota School of Mines and Technology, Rapid City, SD 57701, USA}
\affiliation{Dept. of Physics, University of Wisconsin, River Falls, WI 54022, USA}
\affiliation{Dept. of Physics and Astronomy, University of Rochester, Rochester, NY 14627, USA}
\affiliation{Department of Physics and Astronomy, University of Utah, Salt Lake City, UT 84112, USA}
\affiliation{Oskar Klein Centre and Dept. of Physics, Stockholm University, SE-10691 Stockholm, Sweden}
\affiliation{Dept. of Physics and Astronomy, Stony Brook University, Stony Brook, NY 11794-3800, USA}
\affiliation{Dept. of Physics, Sungkyunkwan University, Suwon 16419, Korea}
\affiliation{Institute of Basic Science, Sungkyunkwan University, Suwon 16419, Korea}
\affiliation{Institute of Physics, Academia Sinica, Taipei, 11529, Taiwan}
\affiliation{Dept. of Physics and Astronomy, University of Alabama, Tuscaloosa, AL 35487, USA}
\affiliation{Dept. of Astronomy and Astrophysics, Pennsylvania State University, University Park, PA 16802, USA}
\affiliation{Dept. of Physics, Pennsylvania State University, University Park, PA 16802, USA}
\affiliation{Dept. of Physics and Astronomy, Uppsala University, Box 516, S-75120 Uppsala, Sweden}
\affiliation{Dept. of Physics, University of Wuppertal, D-42119 Wuppertal, Germany}
\affiliation{DESY, D-15738 Zeuthen, Germany}

\author{R. Abbasi}
\affiliation{Department of Physics, Loyola University Chicago, Chicago, IL 60660, USA}
\author{M. Ackermann}
\affiliation{DESY, D-15738 Zeuthen, Germany}
\author{J. Adams}
\affiliation{Dept. of Physics and Astronomy, University of Canterbury, Private Bag 4800, Christchurch, New Zealand}
\author{J. A. Aguilar}
\affiliation{Universit{\'e} Libre de Bruxelles, Science Faculty CP230, B-1050 Brussels, Belgium}
\author{M. Ahlers}
\affiliation{Niels Bohr Institute, University of Copenhagen, DK-2100 Copenhagen, Denmark}
\author{M. Ahrens}
\affiliation{Oskar Klein Centre and Dept. of Physics, Stockholm University, SE-10691 Stockholm, Sweden}
\author{J.M. Alameddine}
\affiliation{Dept. of Physics, TU Dortmund University, D-44221 Dortmund, Germany}
\author{C. Alispach}
\affiliation{D{\'e}partement de physique nucl{\'e}aire et corpusculaire, Universit{\'e} de Gen{\`e}ve, CH-1211 Gen{\`e}ve, Switzerland}
\author{A. A. Alves Jr.}
\affiliation{Karlsruhe Institute of Technology, Institute for Astroparticle Physics, D-76021 Karlsruhe, Germany }
\author{N. M. Amin}
\affiliation{Bartol Research Institute and Dept. of Physics and Astronomy, University of Delaware, Newark, DE 19716, USA}
% ...
\date{\today}  
\collaboration{IceCube Collaboration} 
% \email{analysis@icecube.wisc.edu}
\noaffiliation

\maketitle

\paragraph{\textbf{Introduction.}}
IceCube is a cubic-kilometer neutrino telescope operating at the South Pole, detecting neutrinos of all flavors with energies from tens of GeV to several PeV, based on the Cherenkov radiation of charged secondaries produced in neutrino interactions in the ice and bedrock \cite{Aartsen:2016nxy}.
In 2013, IceCube discovered high-energy astrophysical neutrinos \cite{1242856} and has more recently found compelling evidence for the gamma-ray blazar TXS 0506+056 being a source of high-energy neutrinos \cite{eaat1378, TXS}. However, the sources responsible for the emission of the majority of the detected {astrophysical} neutrinos are still unknown. Furthermore, several studies imply that the population of gamma-ray blazars detected by the \textit{Fermi} Large Area Telescope (LAT) can only account for a small fraction of the observed astrophysical neutrinos {(e.g., \cite{Murase_Waxman2016, Aartsen_2017, Murase2018_blazar, Neronov2018, Garrappa_2019, Foteini2019, Foteini2019_txs, Yuan2020}).}

Active Galactic Nuclei (AGN) have been considered promising potential sites for high-energy neutrino production, as they are among the most powerful emitters of radiation in the known Universe {\cite{Berezinsky1981, AGNmodel_protheroe1983, AGNmodel_kazanas1986, Begelman1990, Stecker1991, AGNmodel_szabo1994, AGNmodel_mannheim1995, Bednarek1999, AGNmodel_atoyan2001, AGNmodel_mucke2001, AGNmodel_alvarez2004, AGNmodel_Anchordoqui2005}}. They have the potential to accelerate protons up to the highest observed cosmic-ray (CR) energies and are surrounded by high-intensity radiation fields where photo-nuclear reactions with subsequent neutrino production can occur. A common scenario, especially in blazars, {assumes} neutrino production in the relativistic jet. 
However, in this case, the production of neutrinos is boosted due to the relativistic motion of the jet, and therefore dominated by the blazar subclass with their jets closely aligned towards our line-of-sight.

Alternatively, CRs can be accelerated in the core regions of AGN, inside the accretion disk or the corona surrounding it {\cite{Stecker1991, AGNcore_Stecker, Kalashev2015,  Murase+Kimura+Meszaros_2019, Kimura2019, PhysRevD.89.123005, Ali_corona, Inoue2020, Gutierrez2021, Anchordoqui2021, AGNcoronae_review_2021}}. Neutrinos produced in the interactions of these CRs would escape freely, while co-produced GeV and TeV gamma rays would be reprocessed to lower energies in the intense radiation field of the accretion disk before escaping. Indeed, the measured intensity of the extragalactic gamma-ray background (EGB) {\cite{Ackermann_2015, EGB_ackerman2016}} favors such scenarios over gamma-ray transparent neutrino sources \cite{Murase:2016_HiddenSources}.  %\cite{IGRBpaper} 
Production of the observed {astrophysical} neutrino flux in the cores of AGN is compatible with the measured EGB level for both $pp$ \cite{PhysRevD.89.123005} and $p\gamma$ \cite{Stecker1991, AGNcore_Stecker, Kalashev2015} production scenarios.
%Still, given that the non- blazar contribution to the EGB is already constrained below 28% and bright gamma-ray blazars detected by Fermi are disfavoured as the main source of the neutri- nos, a tension remains when considering the gamma-ray flux produced by pγ-neutrino sources. This tension can be resolved when considering photon-photon annihilation inside these sources, caused by the same radiation field as the one responsible for pγ-interactions. As shown in [28], pγ-neutrino sources bright in X-rays or MeV gamma-rays (assuming these can escape the system unhindered) can be such sources. As a result of the many constraints above, conventional sources like gamma-ray bursts and active galactic nuclei bright in gamma rays are unlikely to form the dominant contribution to the diffuse neu- trino flux. Instead, neutrino sources are likely dim in GeV gamma rays [28].In this context, we investigate a model of neutrino sources which are obscured in gamma rays, 

A contribution of AGN cores to the flux of high-energy neutrinos has also been hinted at by the most recent all-sky search for the sources of high-energy neutrinos using 10 years of IceCube data \cite{PS_10year}. The nearby, Seyfert 2 Galaxy NGC 1068 has been found as the most significant point in the sky at a post-trial significance of $2.9\sigma$. The best-fit neutrino flux measured by IceCube for this source exceeds the observed gamma-ray emission by \textit{Fermi}-LAT \cite{Ackermann_2012}, confirming that the sources of high-energy neutrinos could be opaque to high-energy gamma rays in the GeV--TeV range.

In this work, we probe two scenarios of high-energy neutrino production in AGN cores: the first scenario assumes that the neutrino production is proportional to the accretion disk luminosity of the AGN \cite{Stecker1991, AGNcore_Stecker, Kalashev2015}, e.g., due to acceleration of cosmic rays in the disk's corona \cite{Murase+Kimura+Meszaros_2019, Ali_corona}. The accretion disk emits predominantly in the UV {wavelength}. However, a {strong correlation} between UV and X-ray {luminosities} has been observed in AGN, indicating a connection between the primary radiation from the disk and the soft X-ray emission from the hot electrons in the corona \cite{Lusso_2016, Steffen_2006}. Therefore, the soft X-ray flux measured in two all-sky surveys (see below) is used as a proxy for the accretion disk luminosity and the expected neutrino flux in this work.

The second scenario assumes that CR acceleration is suppressed in the AGN with the highest luminosities, favoring instead low-luminosity AGN (LLAGN) for the neutrino production \cite{Kimura_2015, Kimura2019}.
LLAGN are commonly attributed with Radiative Inefficient Acceleration Flows (RIAFs) which are formed instead of an optically thick Shakura-Sunyaev accretion disk \cite{SS73}, when the mass accretion rate into the {super-massive black hole (SMBH)} is relatively small, less than $\approx$~1\% of the Eddington accretion rate. For such low accretion rates, the nuclei in the accretion flow do not thermalize and therefore provide a natural seed population for particle acceleration by, e.g., stochastic processes or magnetic reconnection. The breakdown of this seed population for higher accretion rates would favor particle acceleration in such LLAGN over their more powerful siblings, {and} the neutrino flux would only be proportional to the accretion disk luminosity until the accretion rate limit for RIAFs is reached~\cite{Kimura_2015}. In either scenario, $pp$ interactions and $p\gamma$ interactions can be relevant for generating neutrinos, and gamma rays {would} not escape because of the large optical depths of the accretion disk for photon pair production, thus both models should be regarded as gamma-ray opaque neutrino source models \cite{Kimura_2015}. 

\begin{figure}[t]
    \centering
    \includegraphics[width=0.95\columnwidth]{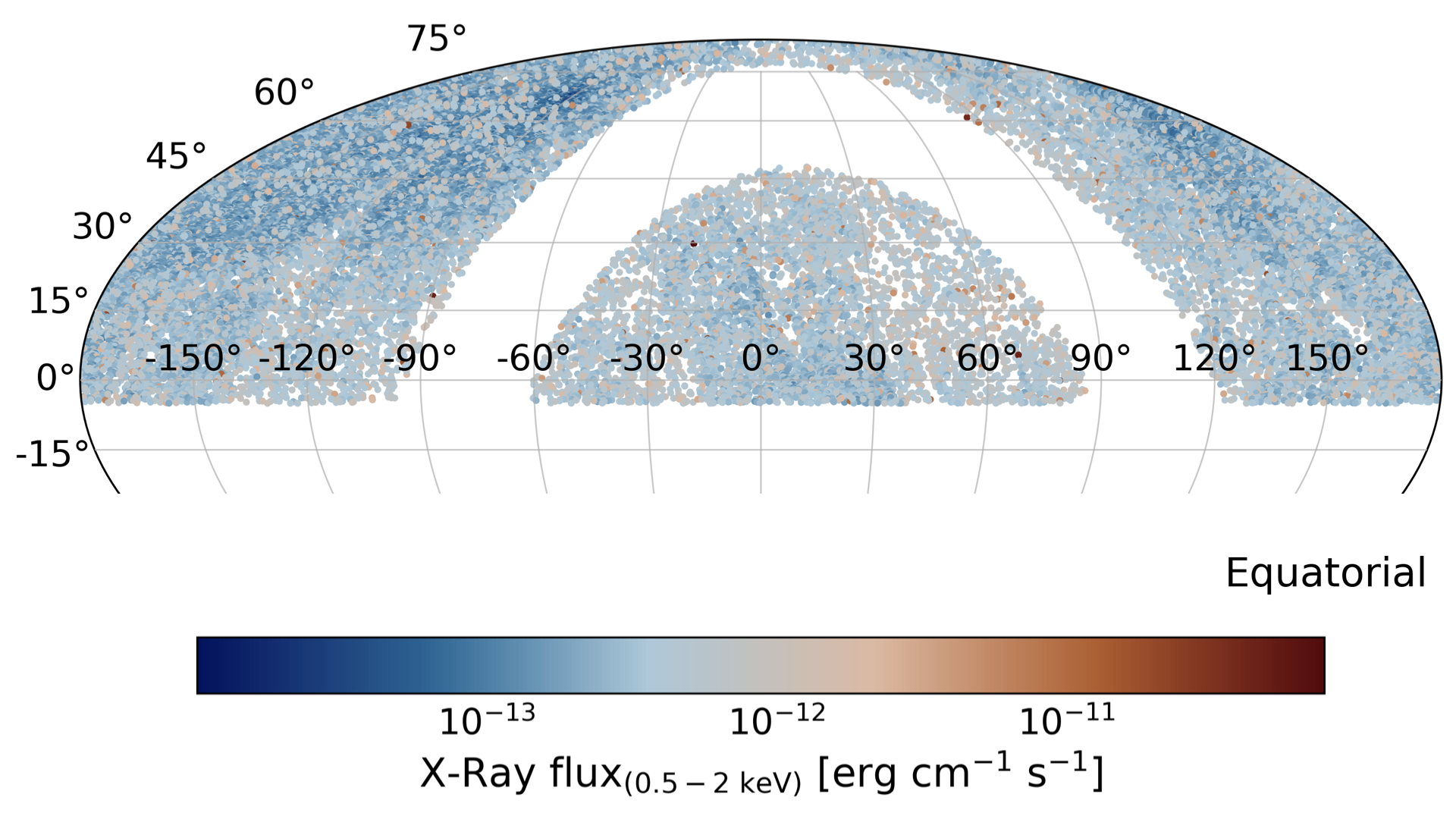}
    \caption{Distribution of sources in the Northern sky ($\delta \ge 5$~deg) for the IR-selected AGN sample (in equatorial Mollweide projection). {Sources in and near the Galactic plane ($|b|> 15^{\circ}$) are excluded from analysis.} The color of the sources show their X-ray flux (2RXS or XMMSL2 X-ray flux, converted in the common soft band 0.5-2~keV \cite{Salvato_2016}) which is used as weight in the analysis.}
    \label{fig:ir_agn}
\end{figure}

\paragraph{\textbf{Active Galactic Nuclei Selection.}}
In order to test the two AGN core models, three samples of AGN have been defined, aiming at a minimal contamination from stellar sources and other galaxy populations. 
The soft X-ray emission is an excellent probe of accretion in AGN, being produced by processes related to the accretion disk. However, based on X-ray data alone, it is challenging to eliminate from the sample the X-ray binaries and galaxies with X-ray emission associated with star formation, rather than an AGN \cite{Hickox:2018xjf}. For this reason, the soft X-ray data is combined with radio and mid-IR wavelengths for the creation of the AGN samples. On one hand, by cross-matching X-ray catalogues with radio catalogues, it is possible to remove non-AGN sources and to selected only luminous AGN, since luminous radio sources are mostly AGN \cite{Miley1980,Condon1992}. Radio emission in AGN is produced by synchrotron emission of electrons, which can be due to processes related to the accretion disk and/or large-scale radio jets \cite{Padovani:2017zpf,Tadhunter_2016}. At the 1.4~GHz, for flux densities of $\sim$10 mJy and above, AGN are strongly dominating over star-forming galaxies \cite{Tadhunter_2016}. Moreover, AGN radio emission is unaffected by dust obscuration, and thus relatively unbiased with respect to orientation. 
On the other hand, the IR wavelength combined with the X-ray data allows to select AGN with emission produced in the core, since the dust surrounding the accretion disk reprocesses the emission of the accretion disk into the IR. The IR waveband is relevant for the identification of many AGN in the Universe that have remained hidden from short-wavelength surveys because of reddening and obscuration by dust in and around their nuclei. This emission dominates the AGN spectrum from wavelengths longer than $\sim 1~\mu$m up to a few tens of microns \cite{Padovani:2017zpf}. It is particularly prominent in the mid-IR (MIR; $3-50~\mu$m) regime, which also contains less stellar contamination. Therefore, IR colors, the ratios of intensities between several mid-IR bands, as collected by WISE~\cite{Wright_2010}, can be used to separate AGN from other source types~\cite{Wright_2010, Salvato_2016}.

Accordingly, we define two samples of luminous AGN, denoted as \emph{radio-selected AGN} and \emph{infrared-selected AGN}. Since they are selected using different criteria (i.e. radio vs IR), they allow to test the same hypothesis (i.e. neutrino emission from the core of luminous AGN) and to make sure the analysis provides a coherent result once it is corrected for the different sky coverage and X-ray contribution of each of the two samples. IR colors can further be used to distinguish between luminous AGN and LLAGN. We use the ratio between the W1 ($\sim3.4~\mu$m) and W2 ($\sim4.6~\mu$m) WISE color bands to define a subset of the infrared-selected AGN that are likely of the low-luminosity type and form our third sample, the \emph{LLAGN}. Additional details about the selection criteria are presented in the \emph{Supplemental Material}.  

\begin{table*}[t!]
    \caption{\label{tab:catalogues}Properties of the AGN samples created for the analysis. The surveys used for the cross-match to derive each sample, the final number of selected sources, % the analysis weighting scheme, 
    cumulative X-ray flux in the 0.5-2~keV energy range from the selected sources {\cite{Salvato_2016}} and the completeness (fraction of total X-ray flux from all AGN in the Universe contained in the sample) are listed.
    }
    \centering
    % \resizebox{\textwidth}{!}{%
    % \setlength{\belowrulesep}{5pt}
    \begin{tabular}{l c c c}
        \toprule
        % & \multicolumn{3}{c}{\textbf{AGN type}} \\
        % \cmidrule[\heavyrulewidth](lr){2-4}
        % & \multicolumn{2}{c}{{AGN}} & {{LLAGN}} \\
        % \cmidrule(lr){2-3}
        & {{Radio--selected AGN}} &  {{IR--selected AGN}} &  {LLAGN}\\
        % \midrule[\heavyrulewidth]
        % \midrule[\heavyrulewidth]
        \midrule
        {Matched catalogues} & \textsc{NVSS} + \textsc{2RXS} + \textsc{XMMSL2} & \textsc{AllWISE} + \textsc{2RXS} + \textsc{XMMSL2} & \textsc{AllWISE} + \textsc{2RXS}\\
        {Nr. of sources} & $9749$ &   $32249$ & $15887$\\
        % \textbf{Weight} & \multicolumn{3}{c}{X-ray flux} \\
        %{Source Weight} & X-ray flux &  X-ray flux & X-ray flux\\
        {Cumulative X-ray flux [erg cm$^{-2}$ s$^{-1}$]} & $7.71\times 10^{-9}$& $1.43\times 10^{-8}$& $7.26\times 10^{-9}$\\
        {Completeness} & $5^{+5}_{-3}\%$ &  $11^{+12}_{-7}\%$ & $6^{+7}_{-4}\%$ \\
        % {Completeness} & $\sim5\%$ &  $\sim11\%$ & $\sim7\%$ \\
        % \midrule[\heavyrulewidth]
        % \multicolumn{4}{l}{\footnotesize$^*$ + Seyfertness PDF cut} \\
        \bottomrule
    \end{tabular}
    % }
\end{table*}

%how the cross-match is performed
The cross-matching between sources in the different {used catalogs} is performed using the \texttt{extcat}  code \footnote{\url{https://github.com/MatteoGiomi/extcats}}. 
The primary X-rays catalogs used are the ROSAT All-sky Survey (2RXS; \cite{2RXS}), and the second release of the XMM-Newton Slew Survey (XMMSL2;  \footnote{\url{https://www.cosmos.esa.int/web/xmm-newton/xmmsl2-ug}}). They have been cross-matched to AllWISE counterparts in \cite{Salvato_2016} and provide 106,573 (17,665) X-ray sources from the 2RXS (XMMSL2) surveys with AllWISE IR counterparts \cite{Wright_2010}, covering $\sim 95\%$ of the extragalactic sky ($|b|> 15^{\circ}$).
The radio-selected AGN sample was compiled by cross-matching the 2RXS and XMMSL2 sources in this catalog with the NRAO VLA Sky Survey (NVSS; \cite{Condon_1998}).
% The LLAGN catalogue is obtained through cross-correlation of the 2RXS and the AllWISE survey.
To avoid biases from the potential neutrino emission of gamma-ray blazars for this analysis, the three {obtained AGN samples} are further cross-matched with the 3LAC \emph{Fermi}-LAT catalogue \cite{3LAC} to remove all known gamma-ray blazars from the final samples. Finally, all sources below a declination of $\delta<-5^{\circ}$ are discarded, as this part of the sky is not covered by the sample of IceCube events used in this analysis, and IceCube's sensitivity {weakens} {rapidly} towards the Southern hemisphere.

\indent The radio-selected AGN sample consists of 9749 sources with an estimated contamination from non-AGN sources of only $\sim 5\%$ and an efficiency of selecting AGN of $\sim 94\%$ (see the \textit{Supplemental Material} for more details{)}. It covers $\sim 55\%$ of the sky. 
The IR-selected AGN sample is the largest sample in this analysis, and consists of 32249 sources shown in \autoref{fig:ir_agn}. The contamination from non-AGN sources here is $\sim 6\%$, for an efficiency of selecting AGN of $\sim 89\%$. 
% It covers $\sim 56\%$ of the sky {[Erik](If all samples have the same sky coverage this could be reduced to one single sentence at the end.)}. 
The LLAGN sample is a subset of the IR-selected AGN sample. A normalized parameter has been defined based on the IR intensity ratios in the WISE W1 and W2 bands, named \emph{Seyfertness}, to distinguish Seyfert-type galaxies which are commonly attributed as LLAGN from their more luminous counterparts (see the \emph{Supplemental Material} for details). Only AGN with a Seyfertness~$\geq 0.5$ are accepted for the LLAGN sample, resulting in a total number of 15887 sources for this sample. All three AGN samples are distributed over $\sim 53\%$ of the sky.

The selection of the sources based on IR color ratios, in particular efficiency, contamination and the Seyfertness parameter, {has} been cross-validated using the 20\% of the sources in the 2RXS catalogue that also have counterparts in the VERONCAT \cite{VERONCAT} catalog, where spectroscopic classifications for each object can be found. 

There is, expectedly, significant overlap between the three AGN samples. About $17\%$ of the IR-selected AGN sources are also found in the radio-selected AGN sample. The LLAGN sample, itself a subset of the IR-selected AGN sample, has about $\sim27\%$ of its sources in common with the radio-selected AGN sample.

\autoref{tab:catalogues} summarizes the properties of the three AGN samples created for this work, including the cumulative X-ray flux from all sources in the respective sample and the \emph{completeness}. 
Completeness is defined here as the ratio between the cumulative X-ray flux included in the sample and the total X-ray flux expected from all AGN in the Universe, estimated using their X-ray luminosity function (luminosity-dependent density evolution model; \cite{Miyaji:1999kt, Ebrero:2008hu, Hasinger:2005sb}). 
The completeness allows an estimation of the contribution from sources not included in the sample, and to extrapolate the analysis results below to the full AGN population. See also the \emph{Supplemental Material} for details on the calculation of the completeness factors.

%%%%%%%%%%%%%%%%%%%%%%%%%%%%%%%%%%%%%%%%%%%%%%%%%%%%%%%%%%%%%%%%%%
\paragraph{\textbf{Analysis.}}
{{A stacking analysis is performed to search for the cumulative signal from each of the defined AGN samples \cite{flarestack_code}, using a neutrino event sample of about 497,000 upward-going neutrinos, collected in eight years of IceCube operations.}} Details about this sample are given in \cite{PS_diffuse8years}. {The sample includes only muon-neutrinos to obtain the necessary pointing accuracy and from declinations  $\delta>-5^{\circ}$ in order to reduce the background of atmospheric muons from cosmic-ray air showers.} %The sample includes only muon-neutrinos from declinations  $\delta>-5^{\circ}$ in order to reduce the background of atmospheric muons from cosmic-ray air showers {and to obtain the necessary pointing accuracy.} % The analysis is performed using the public \texttt{Flarestack} code. \cite{flarestack_code}. 
An unbinned maximum likelihood ratio test is performed, to obtain the best fit for $n_s$, the number of signal events, and $\gamma$, the index of the energy spectrum of the signal events assuming a {single} power-law shape. 
Both a signal and a background PDF enter into the likelihood function (equation 3 and 4 in \cite{PS_diffuse8years}), and are constructed from Monte Carlo simulations as in \cite{PS_diffuse8years}. 

{{In a stacking analysis, the total signal PDF is given by the weighted sum of the signal PDFs for the individual AGN. They enter into the signal PDF weighted with their expected relative contribution to the neutrino flux \cite{Aartsen_2017}.}} %Given the large number of sources in the tested samples, it is essential to weight the signal PDF in the likelihood by the expected neutrino flux from the source.
As described above, the soft X-ray flux reported in the catalogs summarized in \autoref{tab:catalogues} is used as a proxy for the accretion disk luminosity and expected neutrino flux. 
For each of the three AGN samples, the likelihood function is maximized with respect to $n_s$ and $\gamma$.
The log of the likelihood ratio between the best-fit hypothesis and the null hypothesis ($n_s = 0$) forms the test statistic (TS). The obtained TS value is compared to the TS distribution generated entirely by background-only simulations to derive the p-value for the IceCube observations to arise from background fluctuations alone. The p-value is defined as the fraction of the generated background-only samples that has an equal or larger TS value than that obtained from the real data. Simulations with different injected signal strengths are then used to calculate the confidence intervals for $n_s$ and $\gamma$.

\paragraph{\textbf{Results.}}
\autoref{tab:pvalues} shows the obtained TS and the corresponding p-values for the three AGN samples, along with the best-fit parameters obtained from the fit: the number of neutrinos $\hat{n}_s$ and the spectral index $\hat{\gamma}$.  
All  three tested samples show excess relative to background expectations, the largest corresponding to a $1.59 \times 10^{-3}$ ({2.95}$\sigma$) pre-trial p-value for the IR-selected AGN sample. The constraints on the best-fit parameters are shown as profile likelihood contours in \reffig{iragn_llhscan}, where $n_s$ has been converted to the neutrino flux normalization $\Phi$ using the appropriate instrument response functions of IceCube {\cite{PS_diffuse8years}}. 
The pre-trial p-values obtained for the radio-selected and LLAGN samples are 0.03 and 0.08, respectively.

To account for testing three alternative hypotheses, the p-values have to be corrected by trial factors to obtain the final significance of the observation. Since the three AGN samples are not independent but have significant overlap (see \reffig{overlap} in the \textit{Supplemental Material}), simulations are necessary to determine these factors instead of a simple analytic correction.  
In order to correctly take into account the correlation between the three AGN samples, $\mathcal{O}(10^4)$ background-only simulations have been {performed}. For each simulation, three p-values are calculated, one for each AGN sample, and the minimum p-value {{is}} selected. The distribution of the minimal p-values from each background-only simulation is then compared with the pre-trial p-values measured in the analysis. The resulting post-trial p-values are shown in \autoref{tab:pvalues}.
\begin{table}[t]
    \caption{\label{tab:pvalues}P-values and the corresponding significance for the rejection of the background-only hypothesis for the three AGN samples. The values are shown with and without a trial factor correction. The best-fit number of astrophysical neutrino events $\hat{n}_s$ and the best-fit astrophysical power-law spectral index $\hat{\gamma}$ are also shown. }
    \centering
    \resizebox{\columnwidth}{!}{%
    \begin{tabular}{l c c c c}
        \toprule
        % & \multicolumn{3}{c}{\textbf{AGN type}} \\
        % \cmidrule[\heavyrulewidth](lr){2-4}
        % & \multicolumn{2}{c}{{AGN}} & {{LLAGN}} \\
        % \cmidrule(lr){2-3}
         \multirow{2}{*}{{AGN samples}} &  \multirow{2}{*}{{$\hat{n}_s$}} &  \multirow{2}{*}{$\hat{\gamma}$} & Pre-trial p-value  & Post-trial p-value\\
        &  &  & {(significance)} & {(significance)}\\
        % \midrule[\heavyrulewidth]
        % \midrule[\heavyrulewidth]
        \midrule
        % {{Radio--selected AGN}} & 53 & 2.03& $2.84 \times 10^{-2}$ (1.9$\sigma$) & $5.85 \times 10^{-2}$ (1.6$\sigma$)\\
        {{Radio--selected AGN}} & 53 & 2.03& {0.03 (1.90$\sigma$)} & {0.06 (1.58$\sigma$)}\\

        {{IR--selected AGN}} & 105 & 1.94& ${1.59 \times 10^{-3}}$ (${2.95\sigma})$ & ${4.64 \times 10^{-3}}$ ${(2.60\sigma)}$\\
        
        {{LLAGN}} & 35 & 1.96 & {0.08 (1.38$\sigma$)} & {0.16 (${0.99\sigma}$)}\\
        % {{Radio--selected AGN}} & 53 & 2.03& $2.8 \times 10^{-2}$ (1.9$\sigma$) & $5.8 \times 10^{-2}$ (1.6$\sigma$)\\
        % {{IR--selected AGN}} & 105 & 1.94& $1.59 \times 10^{-3}$ ($2.9\sigma$) & $4.5 \times 10^{-3}$ ($2.6\sigma$)\\
        % {{LLAGN}} & 35 & 1.96 & 0.08 (1.4$\sigma$) & $1.6\times10^{-1}$  ($1.0\sigma$)\\
        % \midrule[\heavyrulewidth]
        % \multicolumn{4}{l}{\footnotesize$^*$ + Seyfertness PDF cut} \\
        \bottomrule
    \end{tabular}
     }
\end{table}

% FIG. 2: Ns-gamma 2D likelihood scan
\begin{figure}[t]
    \centering
    \includegraphics[width=0.95\columnwidth]{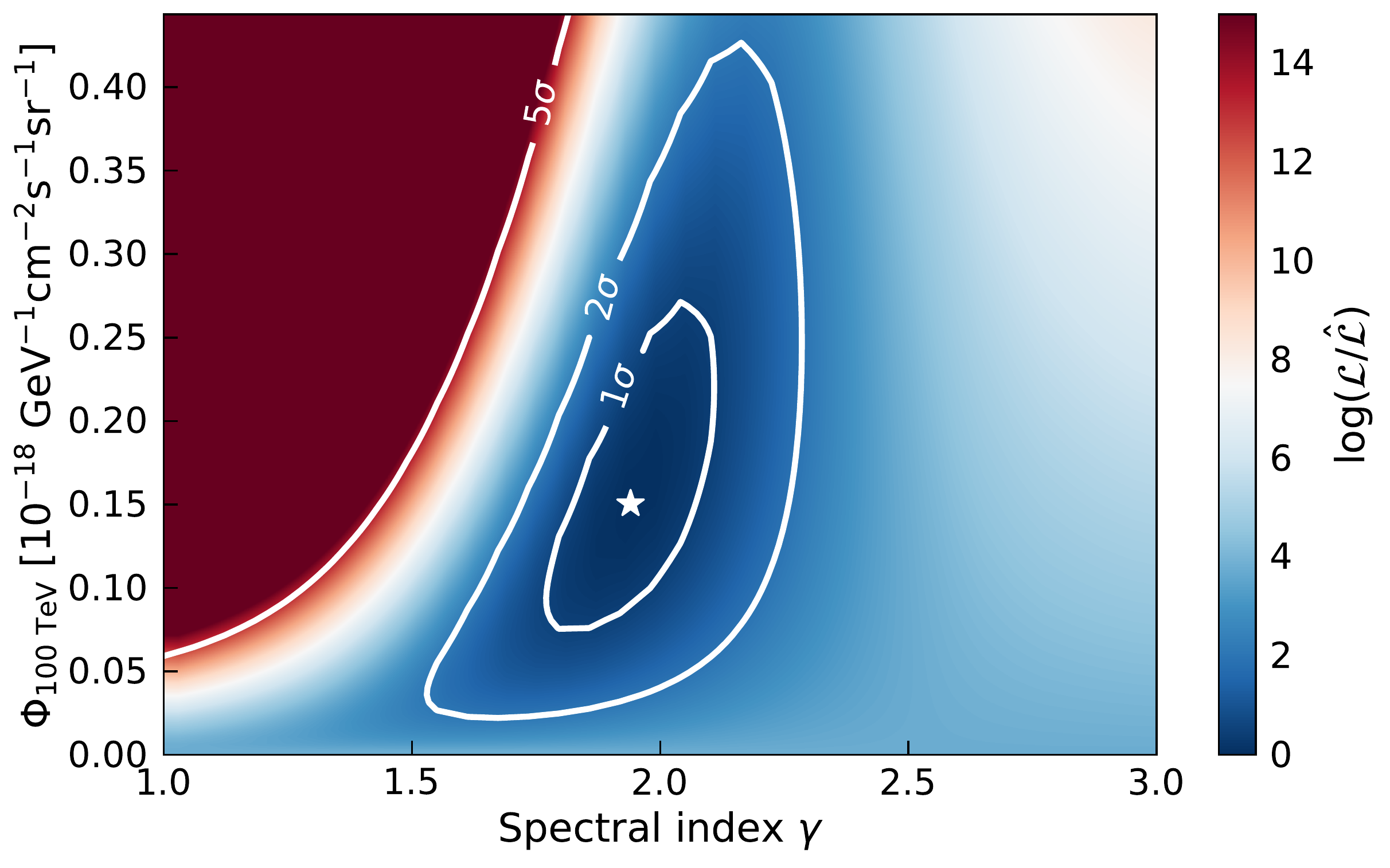}
    \caption{Profile likelihood scans around the best fit for the IR-selected AGN sample as a function of the astrophysical {$(\nu_\mu + \bar{\nu}_\mu)$-flux} spectral index $\gamma$ and normalization $\Phi$ at 100~TeV. Contours show 1, 2 and 5$\sigma$ confidence intervals assuming Wilks' theorem. The best-fit spectral parameters are marked with a star.}
    \labfig{iragn_llhscan}
\end{figure}

The best-fit spectrum for the IR-selected AGN sample, for which the background-only hypothesis is rejected at {2.60}$\sigma$ significance after accounting for multiple trials, is shown in \reffig{best_spectrum} as a quasi-diffuse flux (i.e. the cumulative flux from all objects in the sample divided by $2.2\pi$ to account for the sky coverage), in units of intensity. 
For the other two AGN samples, an excess is also observed but with lower significance. Therefore, upper limits are set for the LLAGN and radio-selected AGN samples, as shown in \reffig{upperlimits}. The upper limits are calculated for an $E^{-2}$  energy spectrum between 30~TeV and 10~PeV, which was found to be the energy range to which this analysis was sensitive.  
For each upper limit and best-fit flux we determine the valid energy range ([30~TeV, 10~PeV]) according to the procedure in \cite{Aartsen_2017} (see the \emph{Supplemental Material} for details). This energy range specifies where IceCube has exclusion power for a particular model. However, given the different spectral index and that the energy range has been calculated based on Monte Carlo simulations, it does not coincide with that of the diffuse flux limits, since in real data no neutrinos with energy greater than few PeV are detected.

The best-fit flux and the upper limits shown in \reffig{best_spectrum} and \reffig{upperlimits} are scaled from the results for the individual samples using the completeness of the catalogues to display the respective values for the total population of AGN. This allows a comparison to the observed astrophysical diffuse neutrino flux measured by IceCube in \cite{globalfit} and \cite{9yrNTdiffuseflux}.  If the result obtained for the IR-selected AGN sample is interpreted as an upper-limit, a neutrino flux up to $1.97 \times 10^{-18}~\mathrm{GeV^{-1}cm^{-2}s^{-1}sr^{-1}}$ is expected at 100~TeV from the cores of the entire population of luminous AGN, assuming an $E^{-2}$ energy spectrum (see the \textit{Summary and Discussion} paragraph for a comparison to the diffuse neutrino flux observed by IceCube).

\begin{figure}[t]
    \centering
        \includegraphics[width=0.95\columnwidth]{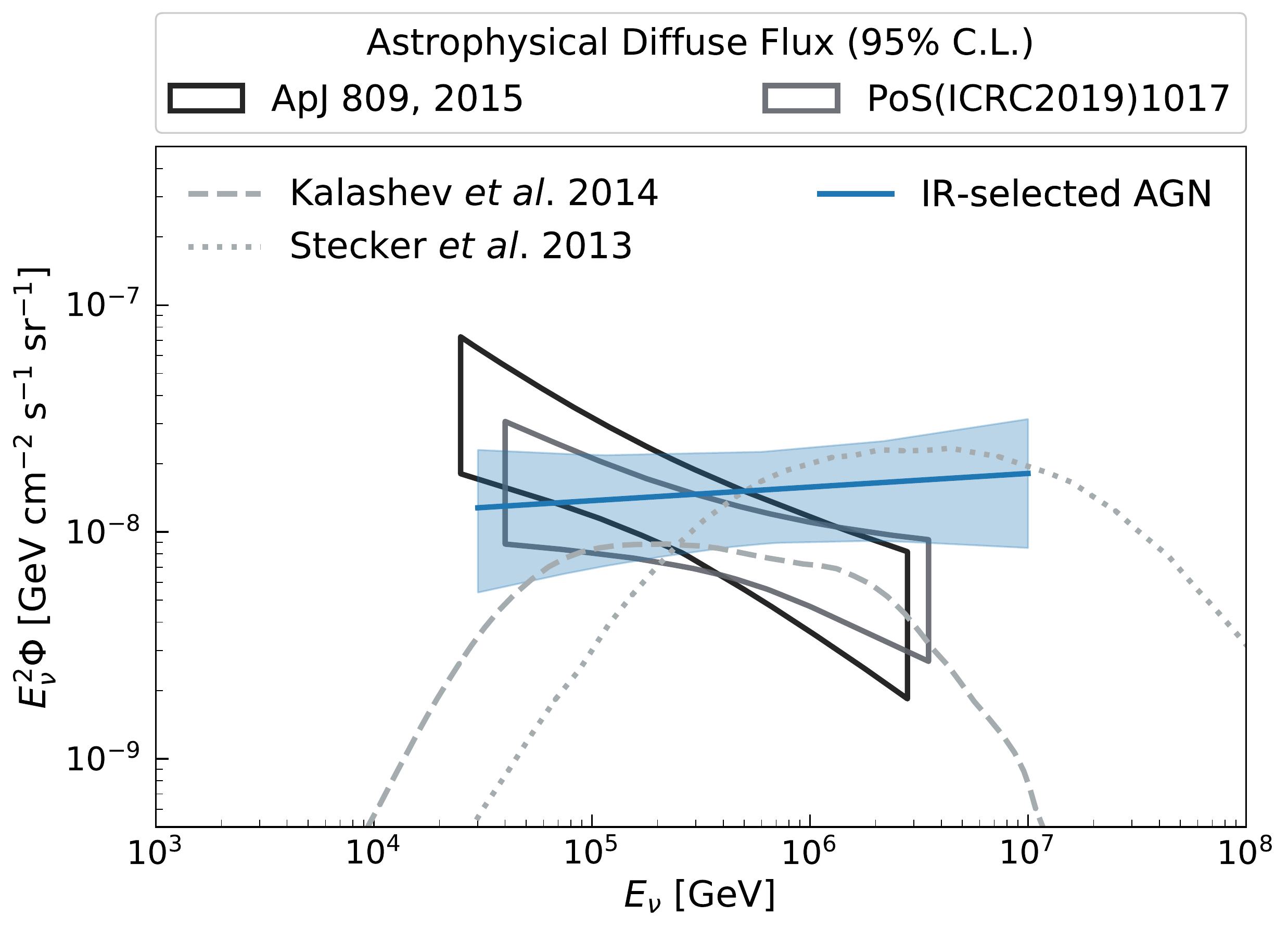}
    \caption{Best-fit astrophysical power-law {$(\nu_\mu + \bar{\nu}_\mu)$-flux} for the IR-selected AGN sample in comparison to the observed astrophysical diffuse neutrino flux. The combined diffuse neutrino flux results from \cite{globalfit} and \cite{9yrNTdiffuseflux} are plotted as a differential flux unfolding using 95\% C.L.  {The best-fit 1$\sigma$ contour is scaled by a correction factor that takes into account the flux from unresolved sources (completeness of the sample). Systematic uncertainties and the error on the completeness factor are not included.} The models from  \cite{Kalashev2015} (dashed, gray line) and \cite{AGNcore_Stecker} (dotted, gray line) are overlaid for comparison.}
    \labfig{best_spectrum}
\end{figure}

% FIG. 4: Differential results for the IR-selected AGN sample with completeness
\begin{figure}[t]
    \centering
    \includegraphics[width=0.95\columnwidth]{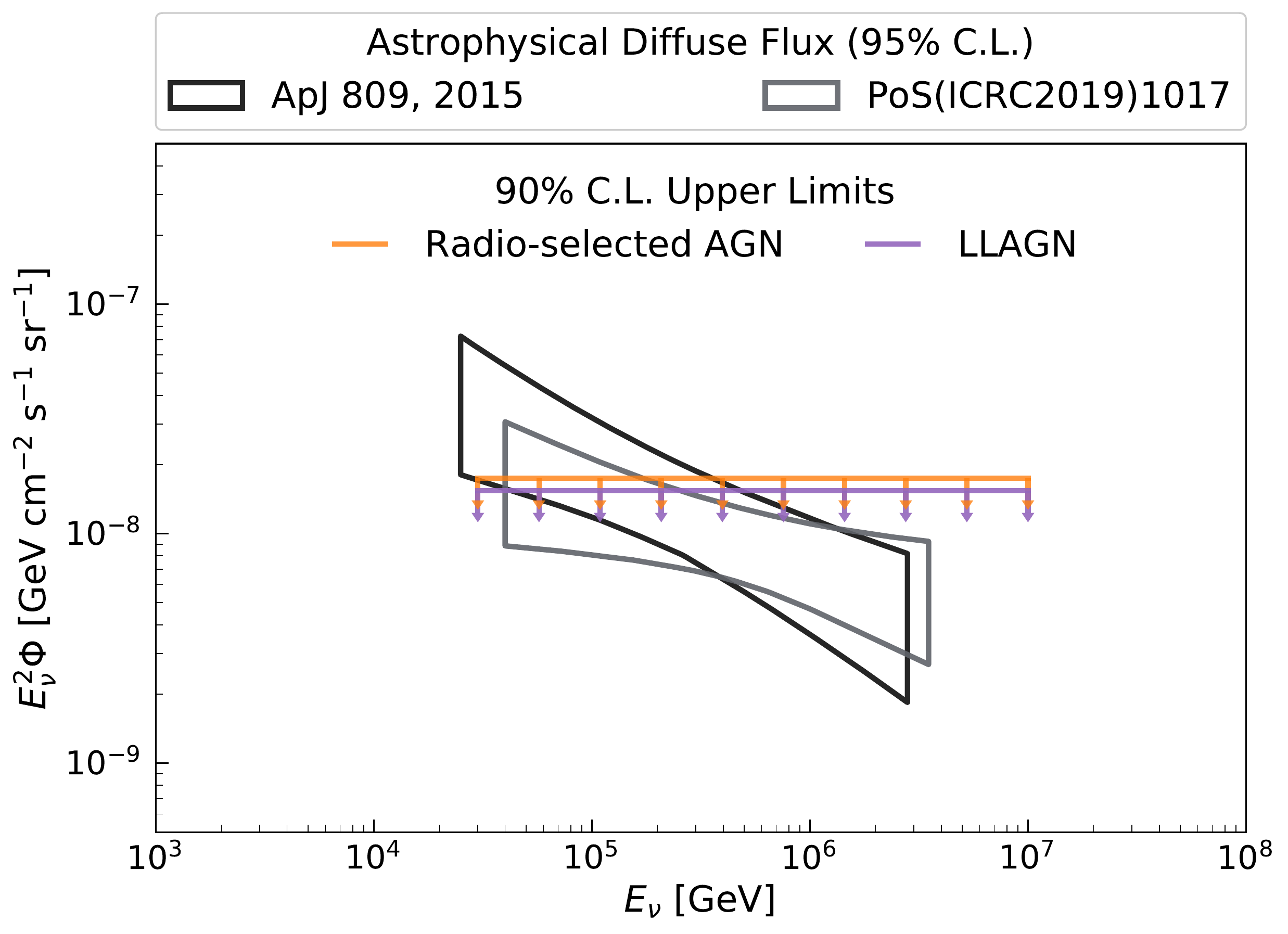}
    \caption{90\% C.L. upper limits on the $(\nu_\mu + \bar{\nu}_\mu)$-flux for the radio-selected AGN and LLAGN populations in comparison to the observed astrophysical diffuse neutrino flux. The combined diffuse neutrino flux results from \cite{globalfit} and \cite{9yrNTdiffuseflux} are plotted as a differential flux unfolding using 95\% C.L. The flux upper limits are shown for a power-law with spectral index 2.0 in the energy range between 30~TeV and 10~PeV. The upper limits include a correction factor that takes into account the flux from unresolved sources (completeness of the samples), while systematic uncertainties are not included {as well as the error on the completeness factor.}}
    \labfig{upperlimits}
\end{figure}

The figures showing best-fit flux and upper limits do not include systematic uncertainties. Following \cite{PS_diffuse8years}, the total systematic uncertainty on the neutrino flux normalization is estimated to be $11\%$. The major contribution to the systematic error comes from uncertainties on the optical efficiency of the {IceCube sensors} and on the optical properties of the Antarctic ice.

\paragraph{\textbf{Summary and Discussion.}}
We have presented an analysis probing the origin of {astrophysical} neutrinos by searching for a correlation between the cores of {AGN} and eight years of IceCube neutrino data. 
Two complementary models for neutrino production have been tested in this paper: one that favors neutrinos to be produced in the geometrically thin, optically thick accretion disks of luminous AGN, and one that predicts the bulk of the neutrino emission from the RIAF of LLAGN. In total, three AGN samples, each one consisting of {$\mathcal{O}(10^4)$} sources, have been {compiled} using radio and IR survey data to identify AGN, and distinguish low-luminosity from high-luminosity objects. {The} soft X-ray flux obtained from the 2RXS and XMMSL2 catalogs is used as a proxy for the accretion disk luminosity and expected neutrino emission. Each one of the (statistically not independent) AGN samples shows a positive correlation to the neutrino data, however for the LLAGN it is weak and compatible with no correlation within $1$ standard {deviation}. The IR-selected AGN sample shows the strongest {indication} for a correlation, with a significance corresponding to ${2.60}$ standard deviations after accounting for trial factors from studying more than one sample. The best-fit spectrum of the correlated events, assuming a power-law shape, has a spectral index close to ${2}$ for all studied samples, as expected for particle acceleration scenarios in cosmic environments, and much harder than the background of atmospheric neutrinos. However, this spectral index is significantly harder than the index seen from IceCube diffuse flux measurements \cite{9yrNTdiffuseflux, globalfit}. This implies that the IceCube diffuse flux might arise from multiple populations of sources with different spectra and that the AGN cores would be responsible for the majority of the emission at the highest energies ($>1$~PeV). {In this scenario, the other populations contributing to the diffuse flux would have softer spectra \cite{5yr_inelasticity, 6yr_cascades, 9yrNTdiffuseflux, 7.5yr_hese, Murase:2016_HiddenSources}.} %{Therefore, the other populations contributing to the diffuse flux must have softer spectra if the AGN cores account for most of the higher energy emission \cite{5yr_inelasticity, 6yr_cascades, 9yrNTdiffuseflux, 7.5yr_hese, Murase:2016_HiddenSources}.}

%\caption{Best-fit astrophysical power-law {$(\nu_\mu + \bar{\nu}_\mu)$-flux} for the IR-selected AGN sample in comparison to the observed astrophysical diffuse neutrino flux. The combined diffuse neutrino flux results from \cite{globalfit} and \cite{9yrNTdiffuseflux} are plotted as a differential flux unfolding using 95\% C.L. The best-fit 1$\sigma$ contours include a correction factor that takes into account the flux from unresolved sources (completeness of the samples), while systematic uncertainties are not included. \textbf{The error on the completeness factor is not included.} The models from  \cite{Kalashev2015} (dashed, gray line) and \cite{AGNcore_Stecker} (dotted, gray line) are overlaid for comparison.}

Within the framework of the tested model, i.e. a linear proportionality between accretion disk luminosity (estimated from soft X-rays) and the neutrino flux, the total contribution of AGN to the astrophysical neutrino flux can be extrapolated using X-ray luminosity functions to estimate the contribution of sources not selected in the source samples.  
{The contribution of the IR-selected AGN themselves to the diffuse flux at 100~TeV measured by IceCube \cite{9yrNTdiffuseflux} amounts to $10^{+5}_{-4}\%$. The associated population’s total contribution can be {{27\% -- 100\%}} after completeness correction, assuming soft X-ray and neutrino luminosities are correlated.} %The minimum contribution of the IR-selected AGN sample to the diffuse flux at 100~TeV measured by IceCube \cite{9yrNTdiffuseflux} amounts to $10^{+5}_{-4}\%$. This fraction can be extended up to $96\pm69\%$ when assuming a correlation between soft X-ray and neutrino luminosity. 
The error on this fraction also includes the error on the completeness, which has been combined with the flux error by a bootstrapping method.
This is consistent with a predominant origin of neutrinos at this energy from the cores of AGN, {while potentially accommodating sub-dominant contributions} from {{blazar}} jets \cite{TXS} and potentially {{tidal disruption events}} \cite{TDE_paper}.
It is also consistent with the contribution extrapolated from the best fit to the radio-selected AGN sample, which tests the same hypothesis, albeit for this sample the correlation is statistically less significant.

Our findings represent the first direct {hint} that cosmic rays accelerated in the AGN core regions are responsible for the bulk of the {astrophysical} neutrino flux observed by IceCube {above 100~TeV. The significance of this finding is {2.60}$\sigma$ post-trial.} % At about {[Erik]$2.6\sigma$ post-trial} significance, the findings described here represent only a first indication that cosmic rays accelerated in the AGN core regions are responsible for the bulk of the neutrino flux. 
Moreover, these results {would} suggest that the sources of high-energy astrophysical neutrinos are dominated by the AGN population and that they should be opaque to GeV--TeV gamma rays. {Several qualitative aspects of the present results}, the hard spectrum of the neutrinos attributed to the signal, the consistency between AGN samples and with the total neutrino flux observed, create an intriguing scenario.
Therefore, it is essential to follow-up on these first indications in the future, using significantly larger IceCube datasets, an improved IceCube angular resolution \cite{ssreco}, the next-generation IceCube-Gen2 detector \cite{IceCubeUpgrade,gen2}, and, potentially, deeper soft X-ray surveys, such as the one currently conducted by eROSITA \cite{eRosita1}.

\paragraph{\textbf{Acknowledgements.}}
The IceCube collaboration acknowledges the significant contributions to this manuscript from Federica Bradascio.
USA {\textendash} U.S. National Science Foundation-Office of Polar Programs,
U.S. National Science Foundation-Physics Division,
U.S. National Science Foundation-EPSCoR,
Wisconsin Alumni Research Foundation,
Center for High Throughput Computing (CHTC) at the University of Wisconsin{\textendash}Madison,
Open Science Grid (OSG),
Extreme Science and Engineering Discovery Environment (XSEDE),
Frontera computing project at the Texas Advanced Computing Center,
U.S. Department of Energy-National Energy Research Scientific Computing Center,
Particle astrophysics research computing center at the University of Maryland,
Institute for Cyber-Enabled Research at Michigan State University,
and Astroparticle physics computational facility at Marquette University;
Belgium {\textendash} Funds for Scientific Research (FRS-FNRS and FWO),
FWO Odysseus and Big Science programmes,
and Belgian Federal Science Policy Office (Belspo);
Germany {\textendash} Bundesministerium f{\"u}r Bildung und Forschung (BMBF),
Deutsche Forschungsgemeinschaft (DFG),
Helmholtz Alliance for Astroparticle Physics (HAP),
Initiative and Networking Fund of the Helmholtz Association,
Deutsches Elektronen Synchrotron (DESY),
and High Performance Computing cluster of the RWTH Aachen;
Sweden {\textendash} Swedish Research Council,
Swedish Polar Research Secretariat,
Swedish National Infrastructure for Computing (SNIC),
and Knut and Alice Wallenberg Foundation;
Australia {\textendash} Australian Research Council;
Canada {\textendash} Natural Sciences and Engineering Research Council of Canada,
Calcul Qu{\'e}bec, Compute Ontario, Canada Foundation for Innovation, WestGrid, and Compute Canada;
Denmark {\textendash} Villum Fonden and Carlsberg Foundation;
New Zealand {\textendash} Marsden Fund;
Japan {\textendash} Japan Society for Promotion of Science (JSPS)
and Institute for Global Prominent Research (IGPR) of Chiba University;
Korea {\textendash} National Research Foundation of Korea (NRF);
Switzerland {\textendash} Swiss National Science Foundation (SNSF);
United Kingdom {\textendash} Department of Physics, University of Oxford.

\bibliography{references}% Produces the bibliography via BibTeX.

% % \clearpage

% % \appendix

\section{SUPPLEMENTAL MATERIAL}
\label{sec:appendix}
\paragraph{\textbf{Radio-selected AGN sample selection.}}
The radio-selected AGN sample is obtained by correlating the NVSS with the  2RXS and the XMMSL2 catalogues, keeping all sources whose radio and X-ray positions differ by less than 60~arcsec. The search radius has been chosen based on the X-ray source positional errors, {which are energy-dependent and on average $\sim 20$~arcsec ($\sim 10$~arcsec) for the 2RXS (XMMSL2) sources \cite{Salvato_2016}}. About 6.12\% (5.93\%) of the 2RXS (XMMSL2) objects have more than one radio counterpart, and for this case the closer NVSS source to the X-ray counterpart is chosen. 

\begin{figure}[t]
    \centering
    \includegraphics[width=0.95\columnwidth]{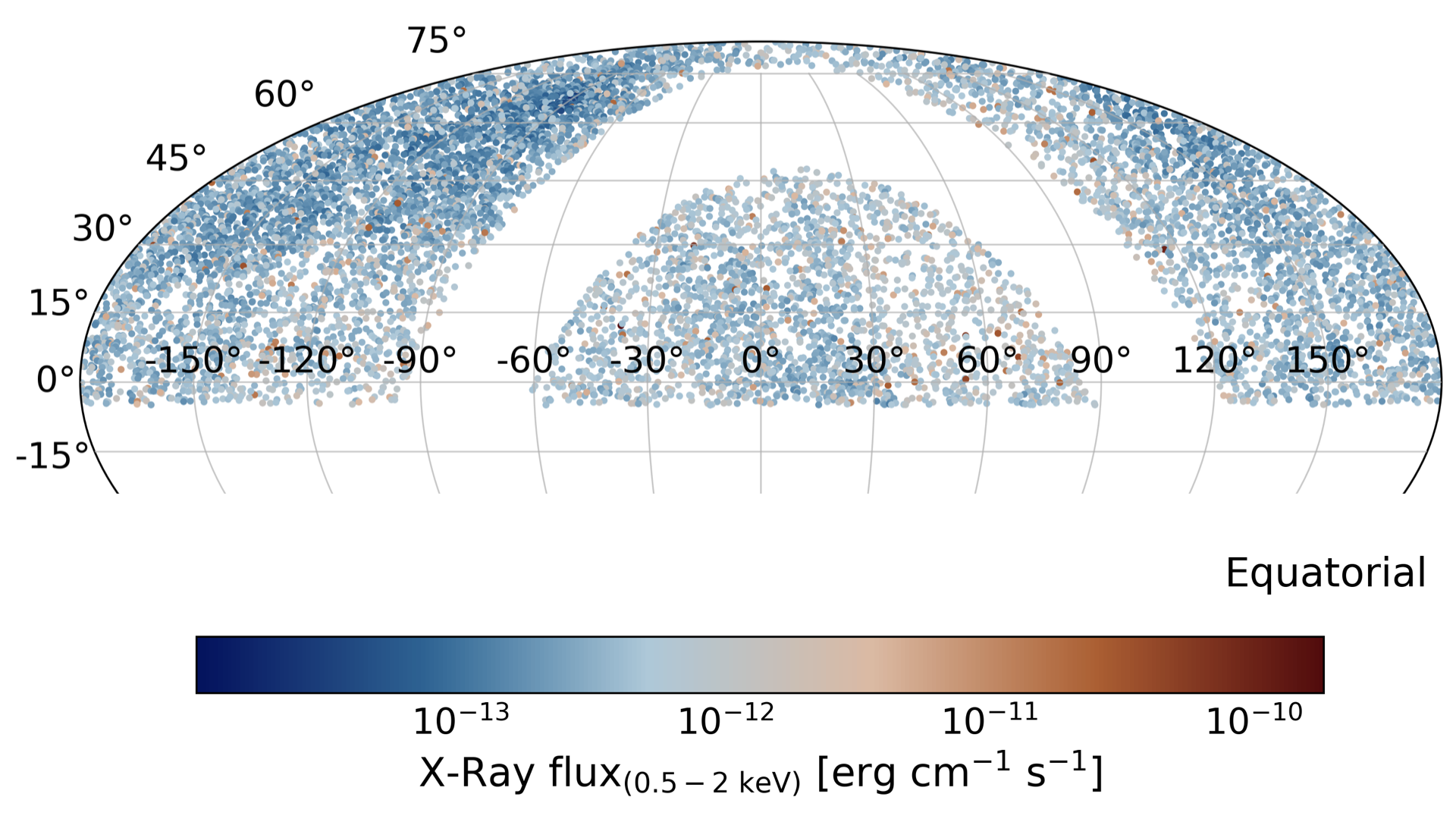}
    \caption{Distribution of sources in the Northern sky ($\delta \ge 5$~deg) for the radio-selected AGN sample (in equatorial Mollweide projection). {Sources in and near the Galactic plane ($|b|> 15^{\circ}$) are excluded from analysis.} The color of the sources show their X-ray flux (in the soft band 0.5-2~keV) which is used as weight in the analysis.}
    \labfig{rlagn}
\end{figure}

\begin{figure}
    \centering
    \includegraphics[width=0.95\columnwidth]{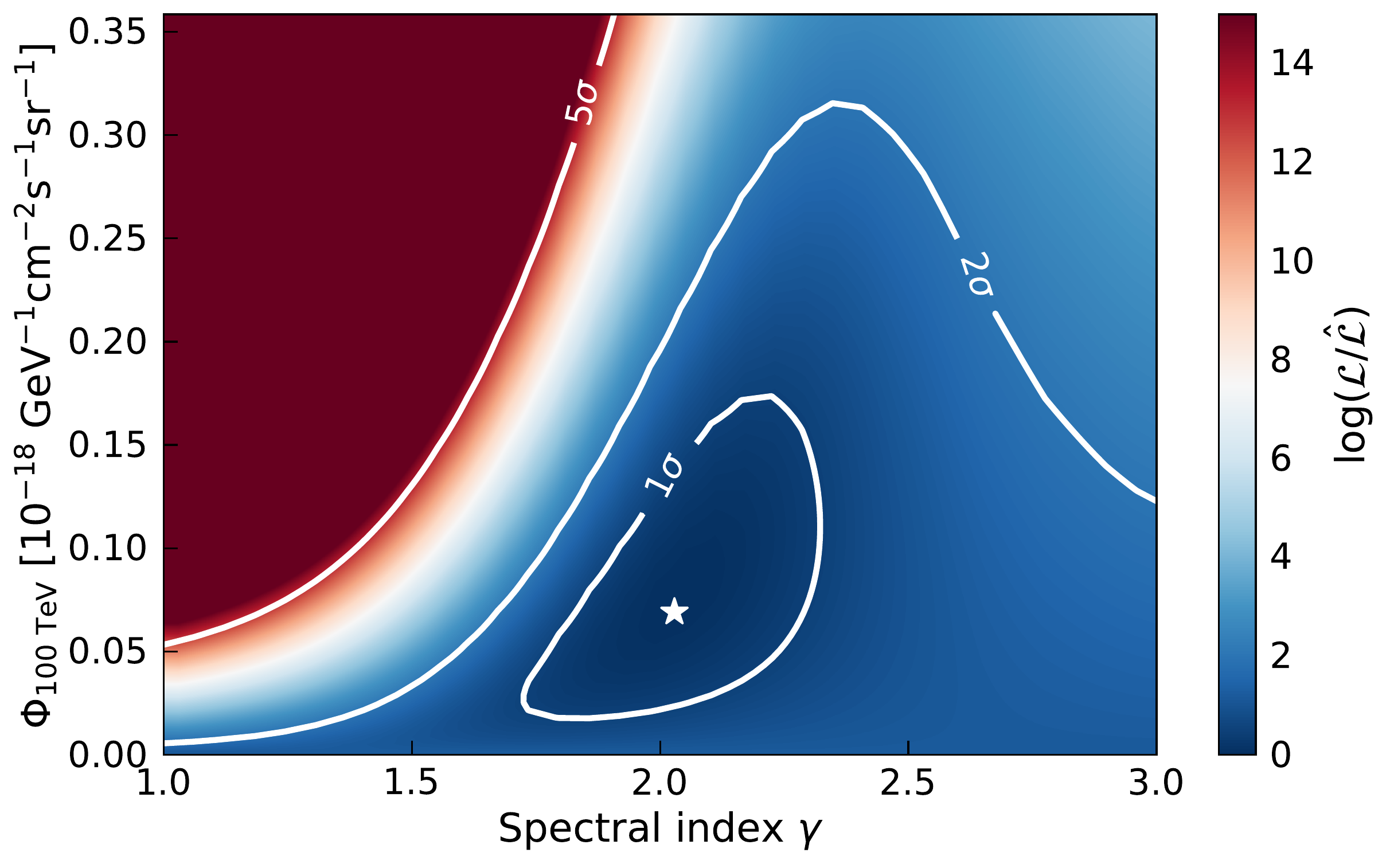}
    \caption{Profile likelihood scans around the best fit for the radio-selected AGN sample as a function of the astrophysical {$(\nu_\mu + \bar{\nu}_\mu)$-flux} spectral index $\gamma$ and normalization $\Phi$ at 100~TeV. Contours show 1, 2 and 5$\sigma$ confidence intervals assuming Wilks’ theorem. The best-fit spectrum is point marked with a star.}
    \labfig{rlagn_llhscan}
\end{figure}
\begin{figure*}[th!]
    \centering
    \includegraphics[width=0.9\textwidth]{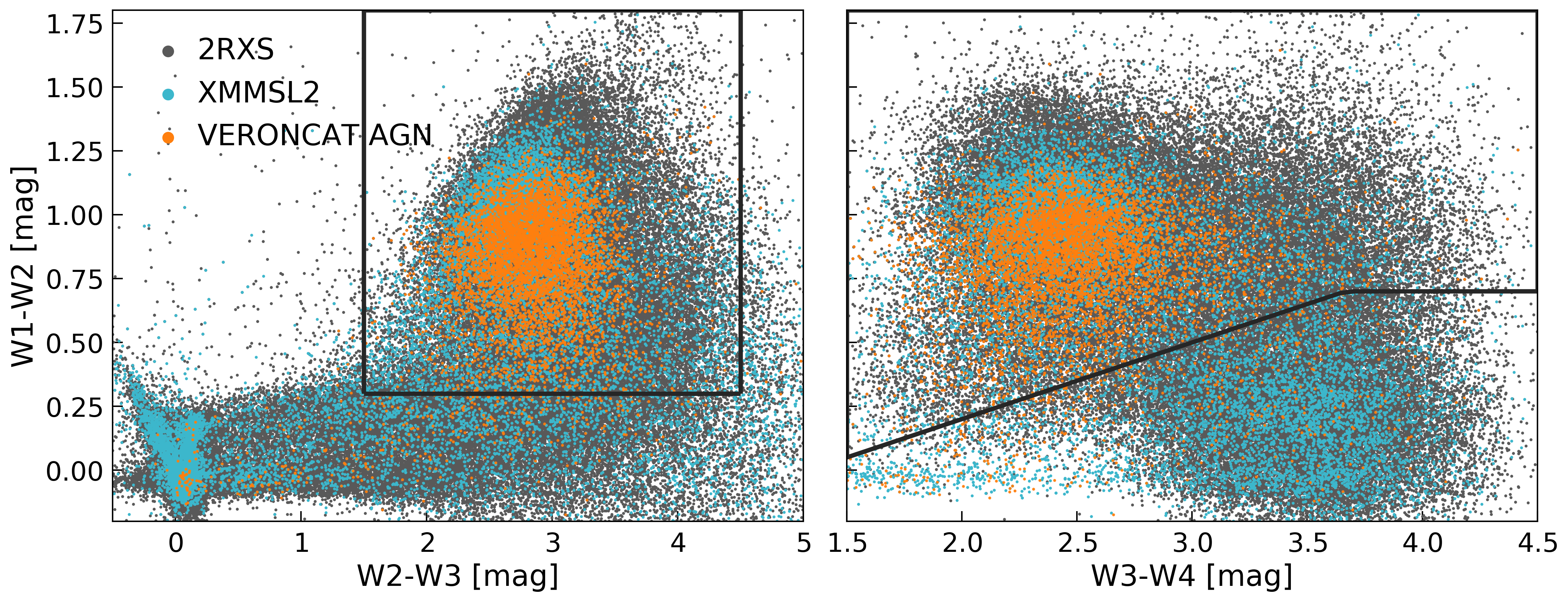}
    \caption{[W1$-$W2] magnitude plotted against the [W2$-$W3] for the AllWISE counterparts to 2RXS (in gray) and XMMSL2 (in light blue). The black lines show the cuts that are going to be applied based on the VERONCAT AGN position (in orange).}
    \labfig{iragn_cuts} 
\end{figure*}

The X-ray sources with a radio counterpart are then cross-matched with the 3LAC catalogue \cite{3LAC} in order to remove blazars, using the 95\% source position error of the gamma-ray sources as search-radius. The largest group of objects in the X-ray/NVSS -- 3LAC correlation are the BL Lacs, followed by FSRQs and blazars of uncertain type. Only a few percent of the sources are non-blazar AGN. 
Before combining the NVSS/2RXS and NVSS/XMMSL2 samples, we remove duplicated X-ray sources (i.e. XMMSL2 sources already included in the 2RXS catalogue) and convert the X-ray fluxes to the common 0.5-2 keV energy range. 
The final radio--selected AGN sample contains 13,927 sources.
A further geometrical cut is applied in the end to select only sources in the Northern hemisphere (declination $>-5$~deg), resulting in the final 9749 sources (see \reffig{rlagn}). {The} contamination from non-AGN sources and efficiency of the selection have also been calculated by cross-matching the NVSS catalogue with two randomized X-ray catalogues reproducing the density distribution of the 2RXS and XMMSL2 catalogues. It results in a contamination of 5\% and an efficiency of $\sim 94\%$.

\reffig{rlagn_llhscan} shows the astrophysical flux normalization $\Phi$ and the spectral indices obtained by minimizing the log-likelihood ratio $\log(L/L_0)$ for the radio-selected AGN sample, assuming an unbroken power-law spectrum.

\begin{figure}
    \centering
    \includegraphics[width=0.9\columnwidth]{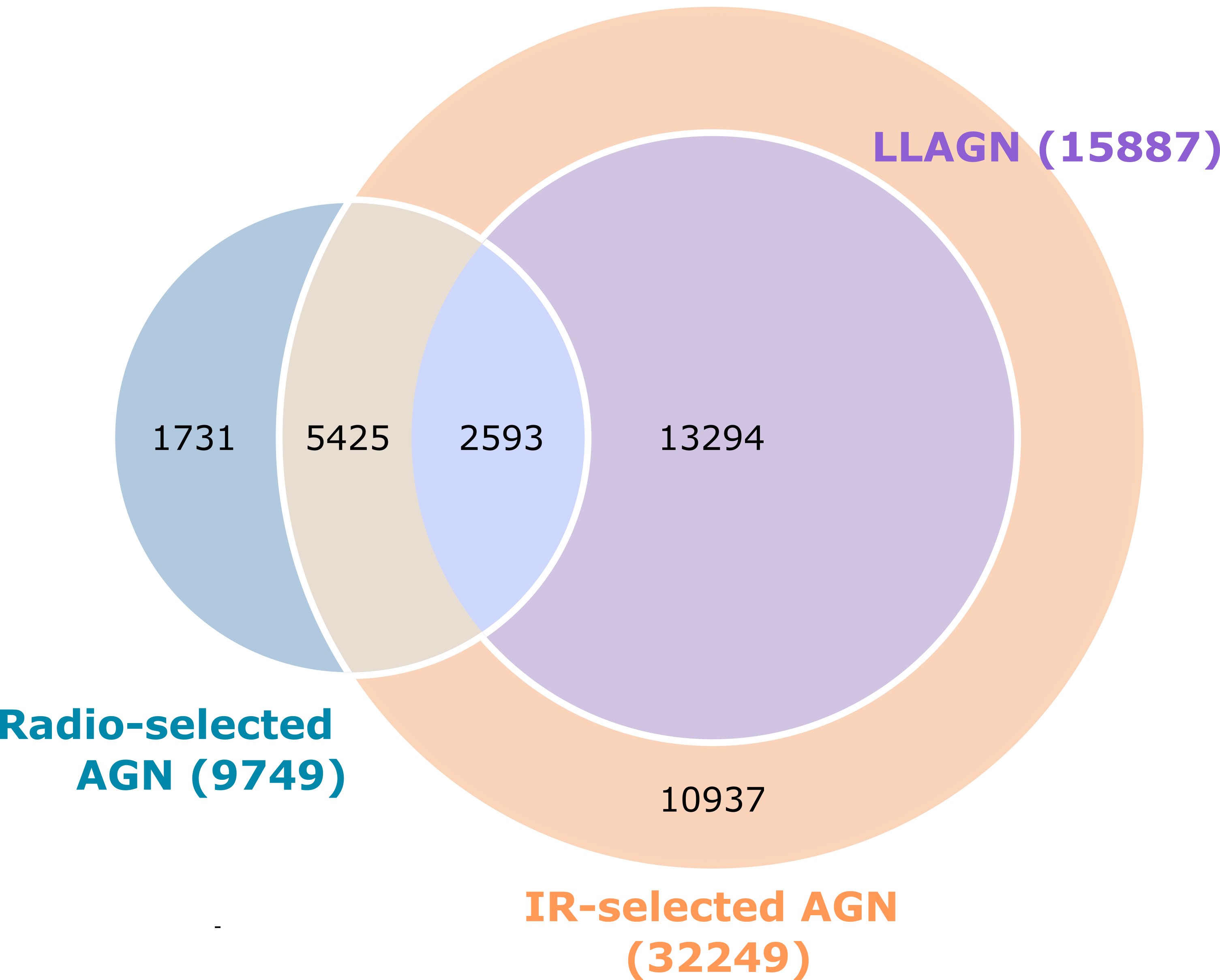}
    \caption{Visualization of the source overlap between the different AGN samples. The number of sources in common is derived via positional cross-match within 60 arcsec search radius. The LLAGN sample is completely included into the IR-selected AGN sample by construction.}
    \labfig{overlap}
\end{figure}

\paragraph{\textbf{IR-selected AGN sample selection.}}
In this paragraph we describe in {detail} the cuts applied on the original 2RXS and XMMSL2 catalogues from \cite{Salvato_2016} to obtain the IR-selected AGN sample. They are based on the following IR-color magnitudes: W1 is defined in the 3.4~$\mu$m band, W2 at 4.6~$\mu$m, W3 and W4 at 12 and 22~$\mu$m, respectively.
The first cut we apply is based on the X-ray/MIR relation suggested by \cite{Salvato_2016}. {It is seen} that the X-ray sources that verify the empirical relation 
\begin{equation}
\label{eq:w1cut}
\mathrm{[W1]}\geq -1.625 \cdot \log F_{(0.5-2\mathrm{keV})} -8.8
\end{equation}
are mainly AGN, which can thus be separated from galaxies and stars over six orders of magnitude. {It is also seen} that most of the sources below the relation are also stars based on their AllWISE colours. 
We confirm the validity of this cut by overlapping the position of the AGN counterparts from the {VERONCAT catalogue \cite{VERONCAT}} to the 2RXS sources in the same plane: most of these AGN verify \autoref{eq:w1cut}, suggesting that most of the sources above this value are indeed AGN. 

After applying the [W1] cut of \autoref{eq:w1cut}, we can use the AllWISE [W1$-$W2] and [W2$-$W3] of the sources for their qualitative characterization as in \cite{Wright_2010} and to isolate the position of the sources we are interested in, removing blazars, Starburst and normal galaxies. Furthermore, we can look at how the VERONCAT counterparts of the X-ray sources {are distributed} in the same color-color diagram to validate these cuts. The VERONCAT counterparts of the 2RXS sources lay in the area in between the black lines in \reffig{iragn_cuts}, that represent the cuts used to isolate the AGN in the [W1$-$W2] and [W2$-$W3] plane. This validates thus the 3 color-color cuts and we can therefore apply them on the 2RXS and XMMSL2 samples (see \reffig{iragn_cuts}, left panel).

The final cut is applied by looking at the [W1$-$W2] vs. [W3$-$W4] plot in the right panel of \reffig{iragn_cuts}. Also in this case, by overlapping the position of the VERONCAT AGN we discover that the sources above the black solid line are AGN, while below that curve the sources are mostly normal galaxies. Therefore, we apply by-eye one more cut to  [W1$-$W2] as a function of [W3$-$W4].

After applying all the color cuts, a cross-match with the 3LAC catalogue is performed to remove the blazars from the final sample.

\begin{figure}[t]
    \centering
    \includegraphics[width=0.95\columnwidth]{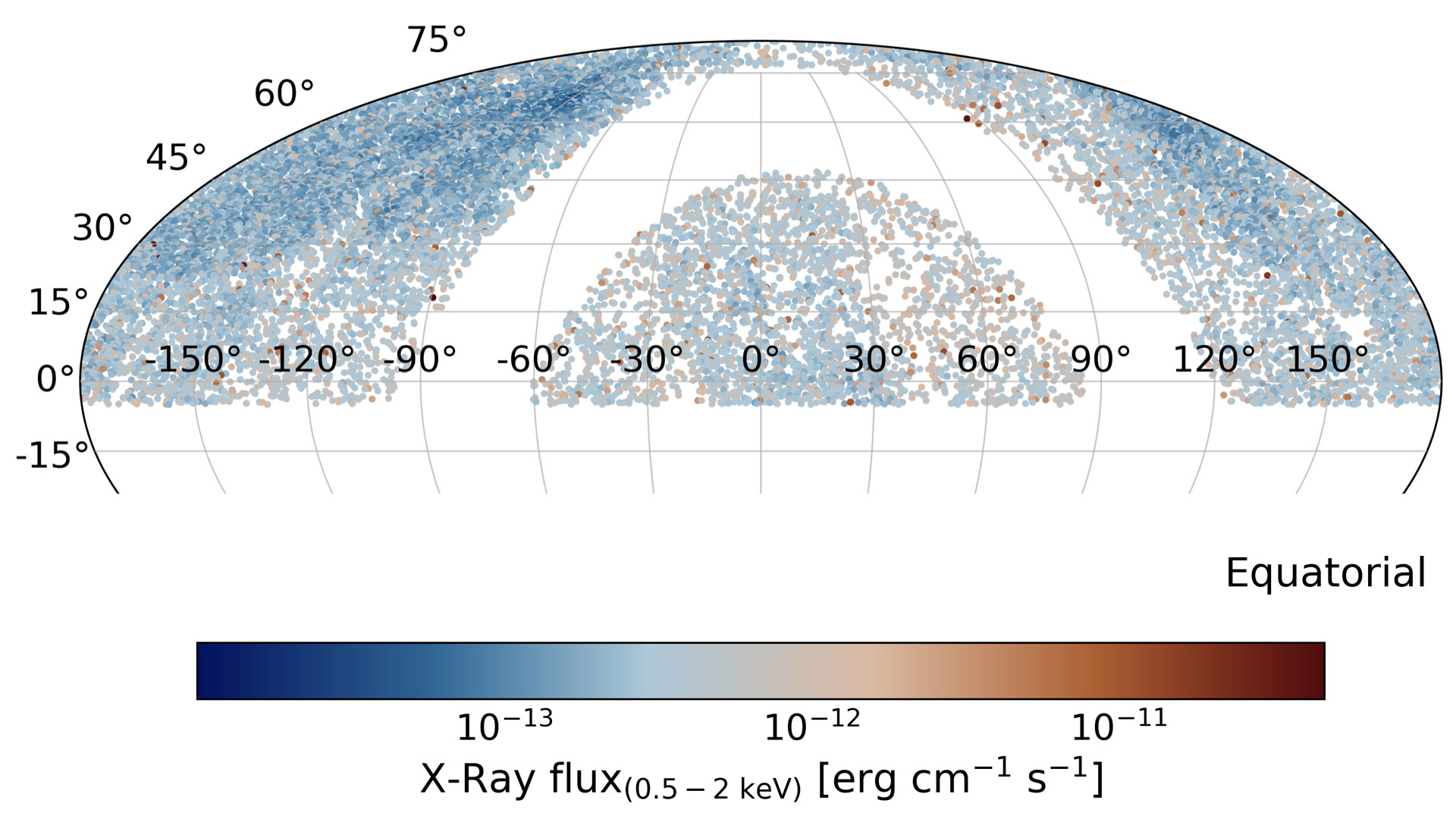}
    \caption{Distribution of sources in the Northern sky ($\delta \ge 5$~deg) for the LLAGN sample (in equatorial Mollweide projection). {Sources in and near the Galactic plane ($|b|> 15^{\circ}$) are excluded from analysis.} The color of the sources show their X-ray flux (in the soft band 0.5-2~keV) which is used as weight in the analysis.}
    \labfig{llagn}
\end{figure}

\begin{figure}[t]
    \centering
    \includegraphics[width=0.95\columnwidth]{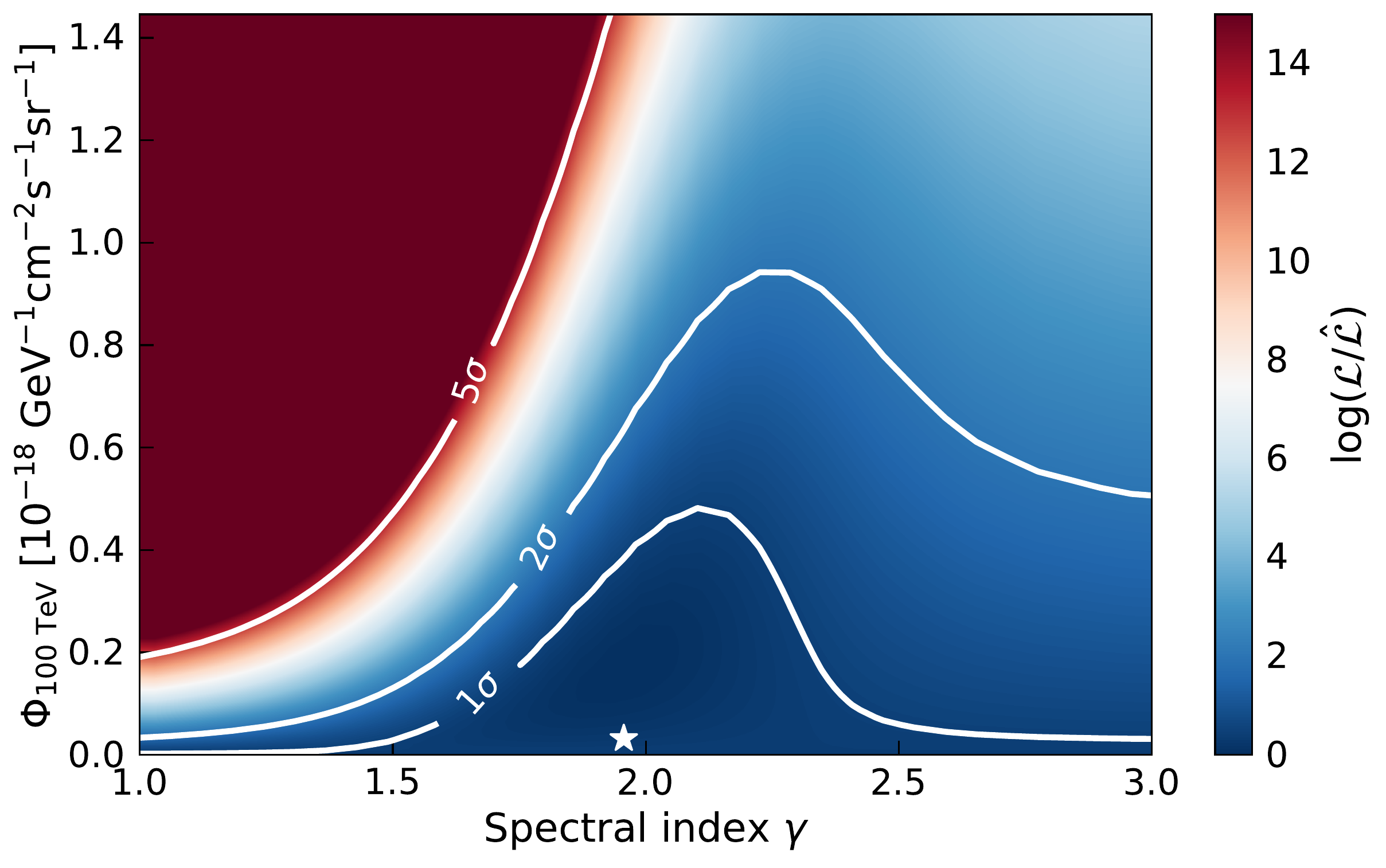}
    \caption{Profile likelihood scans around the best fit for the LLAGN sample as a function of the astrophysical {$(\nu_\mu + \bar{\nu}_\mu)$-flux} spectral index $\gamma$ and normalization $\Phi$ at 100~TeV. Contours show 1, 2 and 5$\sigma$ confidence intervals assuming Wilks’ theorem. The best-fit spectrum is point marked with a star.}
    \labfig{llagn_llhscan}
\end{figure}

\paragraph{\textbf{LLAGN sample selection.}}
The LLAGN sample is a sub-sample of the IR-selected AGN one (see \reffig{overlap}). Only the IR-selected sources with a 2RXS counterpart are selected, since most of them have a counterpart in the VERONCAT catalogue. Using the VERONCAT classification, it is possible to separate the bright AGN from the Seyfert galaxies in the [W1$-$W2] space. We can use these two distributions (after normalizing them) to define a ``Seyfertness''  PDF as:
\begin{equation}
\label{eq:sey_pdf}
    \mathrm{Seyfertness}\ (\mathrm{W1}-\mathrm{W2}) = \frac{P(\mathrm{S})}{P(\mathrm{S})+P(\mathrm{B})}
\end{equation}
where $P(\mathrm{S})$ and $P(\mathrm{B})$ are the probability of being a Seyfert or a bright galaxy, respectively, approximated from the $W1$-$W2$ histograms. Since LLAGN are mostly Seyfert galaxies \cite{doi:10.1146/annurev.astro.45.051806.110546}, we can use this {PDF} to give a weight to our sources based on how likely they are to be LLAGN. We assign to each source a {Seyfertness} PDF between 0 and 1. In the final sample we only include sources with Seyfertness~$\geq 0.5$, since at this value we have the best trade-off between efficiency (77\%) and contamination (21\%) of the selection. The final LLAGN sample contains 25,648 sources and 15887 sources are at a declination larger than $-5$~deg (see \reffig{llagn}).

After applying the stacking analysis to the LLAGN sample, we obtain the 2-D likelihood {profile} shown in \reffig{llagn_llhscan}: the astrophysical flux normalization $\Phi$ is plotted as a function of the spectral indices obtained by the fit assuming an unbroken power-law spectrum.
\begin{figure}[t!]
    \centering
    \includegraphics[width=\columnwidth]{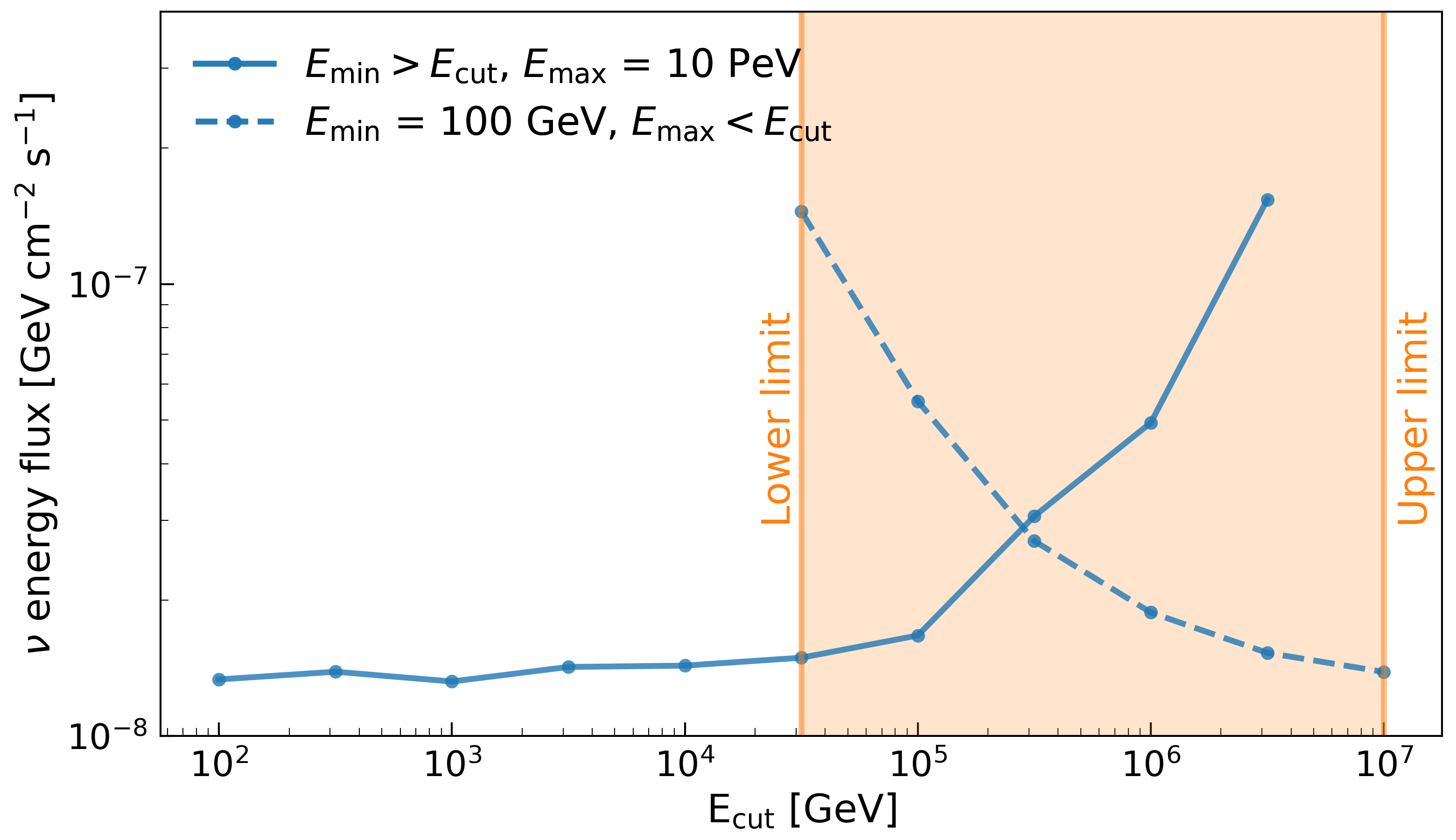}
%    \caption{}
%    \labfig{energy_range2}
%\end{figure*}
%\begin{figure*}[h]
%    \centerin
    \caption{Sensitivity to the integrated neutrino energy flux in the energy range [100 GeV, 10 PeV] as a function of the energy cut for 100 sources of the radio-selected {AGN} sample and $\gamma = 2$. The solid blue line shows the sensitivity when varying $E_{\mathrm{min}}$ bound, while the dashed blue line shows the sensitivity for $E_{\mathrm{min}} = 100~\mathrm{GeV}$ and $E_{\mathrm{max}} = E_{\mathrm{cut}}$.}
    \labfig{energy_range} 
\end{figure}
\paragraph{\textbf{Energy range calculation.}}
\label{sec:energyrange}
The sensitivity  of IceCube depends on the neutrino energy. For this reason, it is important to determine, {for a given spectral} index, the relevant energy range $[E_{\mathrm{min}}, E_{\mathrm{max}}]$ where this analysis is mainly sensitive. Outside this energy range there are few events that contribute little to the {TS}. In order to estimate it, we progressively change the energy range of the injected neutrino signal. {The point at which the sensitivity decreases by 5\% indicates the limit for the analysis.} %The point at which the sensitivity starts deteriorating indicates the limit for the analysis. 
The lower energy bound $E_{\mathrm{min}}$ is determined by varying $E_{\mathrm{min}}$ and only injecting pseudo-signal events with an energy $E > E_{\mathrm{min}}$, while $E_{\mathrm{max}}$ = 10~PeV remains unchanged. Analogously, we determine the upper energy bound by varying $E_{\mathrm{max}}$ and only injecting pseudo-signal events with an energy $E < E_{\mathrm{max}}$, while $E_{\mathrm{min}}$ = 100~GeV remains unchanged.
\reffig{energy_range} shows the results for $\gamma = 2$, using 100 sources of the radio-selected {AGN} sample. The energy range where we are more sensitive is $[30~\mathrm{TeV}, 10~\mathrm{PeV}]$. We expect the same results to be valid also for the other two AGN samples and when stacking more sources.

\begin{figure*}[h!]
    \centering
    \includegraphics[width=0.9\textwidth]{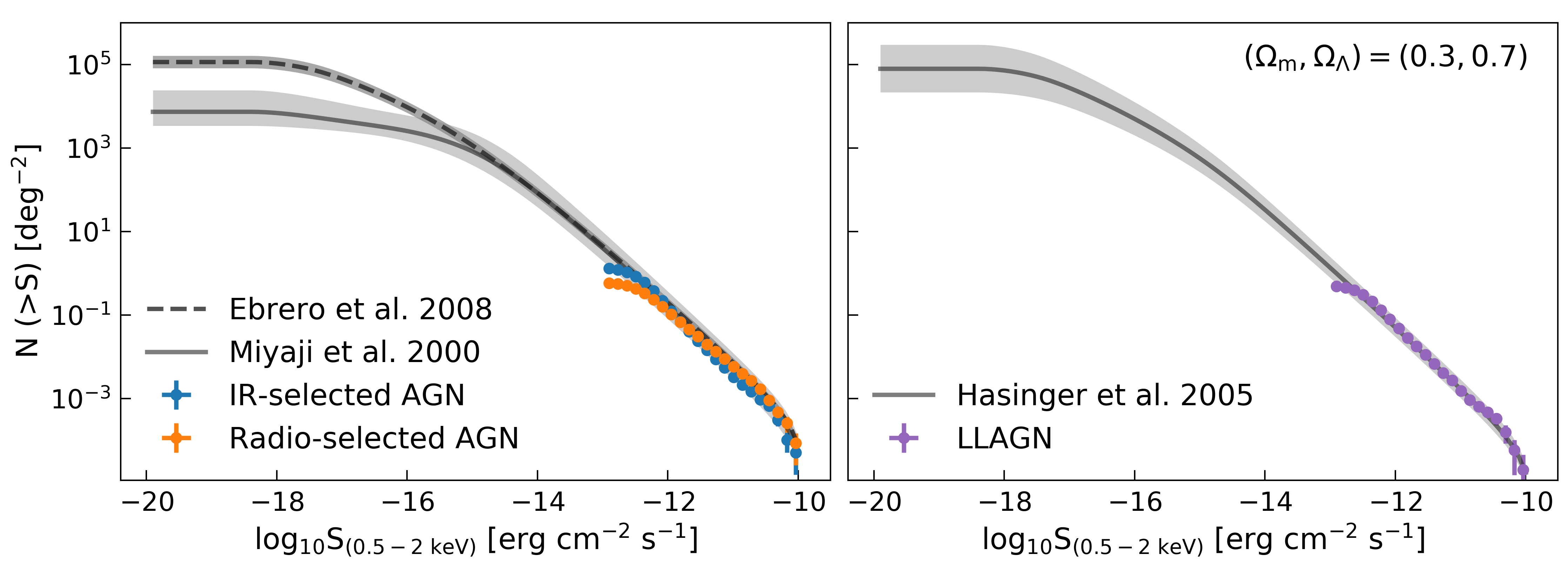}
    \caption{Source count distributions in normalized integral form for sources in the $0.5-2$~keV band. Three models {\cite{Ebrero:2008hu,Miyaji:1999kt,Hasinger:2005sb}} (lines) derived from X-ray luminosity functions are compared with the sources (points) from the three samples created for this analysis. A cosmological framework with $H_0 = 70 \mathrm{\ km \ s}^{-1}\mathrm{\ Mpc}^{-1}$, $\Omega_{\mathrm{M}} = 0.3$ and $\Omega_{\Lambda} = 0.7$ is assumed.
    % Comparison of the normalized integral $0.5 - 2$ keV source count distribution of the catalogues with theoretical models. Error bars correspond to $1\sigma$ confidence. 
    }
    \labfig{lumfunc} 
\end{figure*}

\paragraph{\textbf{Correction factors for unresolved sources.}}
In order to extend the results of the analysis of each sample to the total radio galaxy and LLAGN populations, we need to take into account also those sources not included in our samples. {In fact, some sources can be cut out because of the selections applied for the creation of the samples, either because they are too faint to be detected, or they fall outside the field of view of the X-ray telescopes, or the wavelength of their radiation falls outside the windows investigated here, due to redshift effects.}
%In fact, some sources can be cut out because of the selections applied for the creation of the samples, because they are too faint to be detected, or because they fall outside the field of view of the X-ray telescopes. 
The correction factor is called ``completeness'' and is defined as the fraction of the total flux from all sources in the observable Universe that is resolved into individual point sources in the catalogues used for the analysis. The total X-ray flux expected from all AGN has been estimated using the Soft X-ray Luminosity Function (SXLF) in the energy range 0.5-2 keV. Recent X-ray surveys have found that the SXLF of AGN is best described by a Luminosity-Dependent Density Evolution (LDDE) model, rather than the classical Pure Luminosity Evolution (PLE) or Pure Density Evolution (PDE) models, which tend to overestimate the cosmic X-ray background \cite{Steffen_2006}. According to the LDDE model, the luminosity and the number density distributions change simultaneously as a function of redshift.

By integrating the luminosity functions from \cite{Miyaji:1999kt, Ebrero:2008hu, Hasinger:2005sb} over the luminosity and the comoving volume, we derive the source count distribution $N(>S)$, that is the number of sources detectable at Earth above a minimum photon flux. \reffig{lumfunc} shows the cumulative number of AGN $N(>S)$ as {a} function of the X-ray flux $S$ in the 0.5-2 keV energy range. The expected number of sources derived from the luminosity functions (solid lines) is compared to the number of AGN in the three samples used in this analysis. All samples are scaled by their sky coverage, and the LLAGN sample is also scaled by the efficiency of the selection. 
The completeness factor is obtained by taking the ratio between the area below the three theoretical curves and the area below the distribution of each AGN sample. For the radio-selected and IR-selected AGN samples, the model from \cite{Miyaji:1999kt} is used (left panel of \reffig{lumfunc}), while the LLAGN sample is better described by the model of \cite{Hasinger:2005sb} (right panel of \reffig{lumfunc}). The final values for the completeness are shown in \autoref{tab:catalogues}.

\end{document}